\documentclass[prd,twocolumn,amsmath,amssymb,floatfix,superscriptaddress]{revtex4-1}

\usepackage{graphicx}
\usepackage{amssymb}
\usepackage{amsmath}
\usepackage{hyperref}

\usepackage{color}

\DeclareMathAlphabet\mathbfcal{OMS}{cmsy}{b}{n}
\definecolor{darkgreen}{cmyk}{0.85,0.2,1.00,0.2} 
\definecolor{purple}{cmyk}{0.5,1.0,0,0}

\newcommand{\stucky}{St\"{u}ckelberg}
\newcommand{\parg}{}

\newcommand{\tr}[1]{[#1]}

\newcommand{\ta}{{a}}
\newcommand{\tX}{{X}}
\newcommand{\gisig}{ {\boldsymbol{\Sigma}}}

\def\barray{\begin{array}}
\def\earray{\end{array}}
\def\be{\begin{equation}}
\def\ee{\end{equation}}
\def\ben{\begin{equation} \nonumber}
\def\een{\end{equation}}
\def\ban{\begin{eqnarray*}}
\def\ean{\end{eqnarray*}}
\def\ba{\begin{eqnarray}}
\def\ea{\end{eqnarray}}

\def\({\left(}
\def\){\right)}

\begin{document}

\title{Self-accelerating Massive Gravity:\\ Exact solutions for any isotropic matter distribution}
\author{Pierre Gratia}
\email{pgratia@uchicago.edu}
\affiliation{Department of Physics, University of Chicago, Chicago, Illinois 60637, U.S.A}
\author{Wayne Hu}
\author{Mark Wyman}
\email{markwy@oddjob.uchicago.edu}
\affiliation{Kavli Institute for Cosmological Physics, Department of Astronomy \& Astrophysics,  Enrico Fermi Institute, University of Chicago, Chicago, Illinois 60637, U.S.A}
\begin{abstract}
We present an exact solution to the equations of massive gravity that displays cosmological constant-like behavior for any 
spherically symmetric distribution of matter, including arbitrary time dependence.
On this solution, the new degrees of freedom from the massive graviton generate a cosmological constant-like contribution
to stress-energy that does not interact directly with other matter sources. 
When the effective cosmological constant contribution dominates over other sources
of stress energy the cosmological expansion self-accelerates, even when no other dark-energy-like ingredients are present. The new degrees of freedom introduced
by giving the graviton the mass do not respond to arbitrarily large radial or homogeneous perturbations from other matter
fields on this solution.
 We comment on  possible implications of this result.
\end{abstract}

\maketitle
\section{Introduction}

More than seventy years have elapsed since Pauli and Fierz made the first attempt at writing a theory of gravity with a massive graviton  \cite{Fierz:1939ix}. In the intervening years, daunting challenges to realizing such a theory
have been found, including the scylla of incompatibility with Solar System tests \cite{vanDam:1970vg,Zakharov:1970cc} 
and the charybdis of ghost-like degrees of freedom \cite{Boulware:1972zf}. Recently, de Rham, Gabadadze,
and Tolley  have constructed a theory of massive gravity \cite{Gabadadze:2009ja,deRham:2009rm,deRham:2010ik,deRham:2010kj}  that evades these dangers  \cite{Hassan:2011hr,Hassan:2011ea}. This theory also contains a vacuum solution that recovers exactly a Schwarzschild-de Sitter solution \cite{Koyama:2011xz,Koyama:2011yg}.  Moreover, in the flat matter dominated limit, the theory has a solution that
 responds to the presence of matter by producing an effective cosmological constant contribution to the stress tensor 
 at the cost of introducing inhomogeneous 
solutions for the \stucky\ fields that describe the new degrees of freedom that come from massive gravity
 \cite{D'Amico:2011jj}.  For an open universe, a related solution has been explicitly shown 
 to evolve into self-acceleration \cite{Gumrukcuoglu:2011ew}.
  
 In this paper, we generalize considerations  in \cite{D'Amico:2011jj,Gumrukcuoglu:2011ew} to an arbitrary spatially isotropic metric.   We find cosmological constant type solutions  in the presence of any isotropic distribution of matter. 
Such solutions connect the flat matter dominated solution \cite{D'Amico:2011jj} 
to the de Sitter solution \cite{Koyama:2011xz,Koyama:2011yg} allowing a cosmological expansion history identical
to the  $\Lambda$CDM model even in the presence of spherically symmetric matter perturbations.

\section{Massive Gravity}

The covariant Lagrangian density for a theory of massive gravity will have, in addition to the usual Einstein-Hilbert term, a mass term
with a 
potential $\mathcal U$,
\begin{align}
\mathcal{L}_G &=\frac{M_{\rm pl}^2}{2}\sqrt{-g}\left[ R-\frac{m^2}{4}\mathcal{U}(g_{\mu\nu},\mathcal{K}_{\mu\nu})\right].
\end{align}
$M_{\rm pl}^2 = 1/8\pi G$ and $\hbar=c=1$ throughout.
 $\mathcal{K}_{\mu\nu}$ is a tensor that characterizes metric fluctuations away from a fiducial (flat) space time.
At the linearized level, the potential must take on the Fierz-Pauli structure to be ghost free; but any purely linear theory 
 exhibits the vDVZ discontinuity \cite{vanDam:1970vg,Zakharov:1970cc}, where an extra helicity mode couples to matter even in the $m \rightarrow 0$ limit.
Nonlinear extensions to the Fierz-Pauli potential can evade this problem via
 a strong coupling phenomenon known as the Vainshtein mechanism \cite{Vainshtein:1972sx},
where the extra coupling is suppressed near matter sources. However, these extensions
typically contain an unhealthy ghost-like degree of freedom \cite{Boulware:1972zf}.    

For a 
theory of massive gravity to be free from this ghost, the potential term must take a special form built out of expressions that 
have the form of total derivatives in absence of dynamics \cite{deRham:2010kj}. 
These can be written as contractions of the tensor
\begin{equation}
\mathcal{K}^{\mu}_{\;\,\nu}=\delta^{\mu}_{\;\,\nu}
- \sqrt{\gisig}^{\mu}_{\;\,\nu}.
\label{eqn:kappa}
\end{equation}
The matrix $\sqrt{\gisig}$ is understood to denote
$\sqrt{\gisig}^{\mu}_{\;\,\alpha} \sqrt{\gisig}^{\alpha}_{\;\,\nu} \equiv 
\Sigma^{\mu}_{\;\,\nu}$.
The potential-generating matrix is defined as
\begin{equation}
\Sigma^{\mu}_{\;\,\nu} \equiv g^{\mu\alpha}\partial_{\alpha}\phi^a\partial_{\nu}\phi^b\eta_{ab}\equiv g^{\mu\alpha}\Sigma_{\alpha\nu},
\label{eqn:potentialmatrix}
\end{equation}
where  $\phi^a$ are the 4 \stucky\ fields introduced to restore diffeomorphism invariance.  The $\phi^a$ fields transform as scalars,
while $\Sigma$, $\sqrt{\gisig}$ and $\mathcal{K}$
transform as  tensors under general coordinate transforms. 

In matrix notation, the potential can be written \cite{deRham:2010ik,deRham:2010kj}
\begin{align}
-{\mathcal U / 4} 
&=\tr{\mathbfcal{K}}^2
 -  \tr{\mathbfcal{K}^2} 
+ \alpha_3
\left( \tr{\mathbfcal{K}}^{3} - 3 \tr{\mathbfcal{K}}\tr{\mathbfcal{K}^2}
+ 2\tr{\mathbfcal{K}^{3}}\right)
\nonumber\\ & \quad + \alpha_4
\big( \tr{\mathbfcal{K}}^{4} - 6 \tr{\mathbfcal{K}}^2 \tr{\mathbfcal{K}^2}
+ 8 \tr{\mathbfcal{K}} \tr{\mathbfcal{K}^{3}}
+ 3  \tr{\mathbfcal{K}^2}^2
\nonumber\\& \qquad 
 - 6 \tr{\mathbfcal{K}^{4}}
\big),
\end{align}
where brackets denote traces, $\tr{{\bf A}} \equiv A^\mu_{\;\,\mu}$, and $\alpha_3$, $\alpha_4$ are 
free parameters.
Using Eq.~(\ref{eqn:kappa}), we can reexpress the potential in 
terms of traces of products of $\sqrt{\gisig}$
\begin{align}
{\mathcal U \over 4} &=
-12+6\tr{\sqrt{\gisig}}+\tr{\gisig}-\tr{\sqrt{\gisig}}^2 
\nonumber\\&\quad
+\alpha_3 \Big( {-24} + 18\tr{\sqrt{\gisig}} - 6\tr{\sqrt{\gisig}}^2 + \tr{\sqrt{\gisig}}^3
\nonumber\\&\qquad
- 3 \tr{\gisig} (\tr{\sqrt{\gisig}}-2) + 2 \tr{\gisig^{3/2}}\Big)
\nonumber\\&\quad
+\alpha_4 \Big( {-24}+  24 \tr{\sqrt{\gisig}} -12 \tr{\sqrt{\gisig}}^2- 12 \tr{\sqrt{\gisig}} \tr{\gisig} 
\nonumber\\&\qquad
+ 6\tr{\sqrt{\gisig}}^2 \tr{\gisig}+ 4\tr{\sqrt{\gisig}}^3 + 12  \tr{\gisig} - 3 \tr{\gisig}^2\nonumber\\&\qquad
 - 8  \tr{\gisig^{3/2}}
(\tr{\sqrt{\gisig}}-1)
+ 6\tr{\gisig^2} -  \tr{\sqrt{\gisig}}^4 \Big).
\label{eqn:pottrace}
\end{align}

Variation of the action with respect to the metric yields the modified Einstein
equations
\begin{equation}
G_{\mu\nu} =  m^2 T_{\mu\nu}^{(\mathcal{K})} + \frac{1}{M_{\rm pl}^2} T_{\mu\nu}^{(m)},
\label{eqn:modeinstein}
\end{equation}
where $G_{\mu\nu}$ is the usual Einstein tensor and $T_{\mu\nu}^{(m)}$ is the matter stress energy tensor.  Here 
\begin{align}
T^{(\mathcal{K})}_{\mu \nu}& =  \frac{1}{\sqrt{-g}} \frac{\delta}{\delta g^{\mu \nu}}
\sqrt{-g} \, \frac{\mathcal U } {4} \\
 = &-\frac{1}{2} 
 \Bigl\{ \frac{ \mathcal{U}} {4} g_{\mu \nu}
   - 2 \Sigma_{\mu \nu} -2(3-\tr{\sqrt{\gisig}})\sqrt{\gisig}_{\mu\nu}
   \nonumber\\& 
  +\alpha_3 \Big[
    -  3 \(6 - 4 \tr{\sqrt{\gisig}} +  \tr{\sqrt{\gisig}}^2 -  \tr \gisig \)  \sqrt{\gisig}_{\mu\nu} 
       \nonumber\\& \quad
    + 6 \( \tr{\sqrt{\gisig}} -2\) \Sigma_{\mu \nu} 
    - 6  \gisig^{3/2}_{\mu\nu} \Big]
   \nonumber \\
      &+ \alpha_4 \Big [ - 24 \( \gisig_{\mu \nu}^2 -  ( \tr{\sqrt{\gisig}} -1 ) \gisig_{\mu \nu}^{3/2}  \)  \nonumber \\&\quad
   - 12 \( 2 - 2 \tr{\sqrt{\gisig}} - \tr{\gisig} + \tr{\sqrt{\gisig}}^2
     \) \Sigma_{\mu \nu}   \nonumber  \\
&\quad -   \(24   - 24 \tr{\sqrt{\gisig}} + 12  \tr{\sqrt{\gisig}}^2 - 4\tr{\sqrt{\gisig}}^3 \right . \nonumber \\
&\quad \left .- 12 \tr{\gisig} + 12 \tr{\gisig} \tr{\sqrt{\gisig}} - 8 \tr{\gisig^{3/2}} \) 
\sqrt{\gisig}_{\mu \nu} \Big ]     \Bigr\} \nonumber
\end{align}
is the dimensionless effective stress energy tensor provided by the mass term.
Note that this effective stress energy depends explicitly on the metric itself.
To solve the modified Einstein equation, we first parameterize the metric and then solve for the joint effect of the matter and mass term.

\section{Exact solution}

Generalizing \cite{D'Amico:2011jj, Gumrukcuoglu:2011ew}, we consider an arbitrary spatially isotropic metric,
\begin{equation}
ds^2=-b^2(r,t) dt^2+\ta^2(r,t)(dr^2+r^2d\Omega^2).
\label{eqn:metric}
\end{equation}
We correspondingly take a spherically symmetric ansatz for the \stucky\ fields:
\begin{align}
\phi^0 &= f(t,r) ,\nonumber\\
\phi^i &= g(t,r) \frac{x^i}{r} ,
\end{align}
and look for solutions to the functions $g(t,r)$ and $f(t,r)$.  The potential matrix (\ref{eqn:potentialmatrix}) then takes the form
\begin{equation}
\gisig = \left(
      \begin{array}{cccc}
        \dfrac{\dot{f}^2-\dot{g}^2}{b^2} & \dfrac{\dot{f}f'-\dot{g}g'}{b^2} & 0 & 0 \\
        \dfrac{\dot{g}g'-\dot{f}f'}{\ta ^2} & \dfrac{-f'^2+g'^2}{\ta ^2} & 0 & 0\\
        0 & 0 & \dfrac{g^2}{\ta ^2r^2} & 0\\
        0 & 0 & 0 & \dfrac{g^2}{\ta ^2r^2}
      \end{array} \right),
  \end{equation}
  where primes denote derivatives with respect to $r$ and overdots with respect to $t$.

The resulting, rather involved, calculation is made easier by isolating the upper-left-hand $2 \times 2$
submatrix of $\gisig$ and using the Cayley-Hamilton theorem, which states that a matrix 
solves its own characteristic polynomial. For a $2\times 2$ matrix ${\bf A}$, this means
$$
 \tr{{\bf A}} {\bf A} = {\bf A}^2 + (\det{{\bf A}}) \;  {\bf I}_2,
$$
where ${\bf I}_2$ is the $2 \times 2$ identity matrix. We can then use that $\det {\bf A}^n  = \(\det {\bf A}\)^n$
to find the square root of $\gisig_2$, the upper-left-hand $2 \times 2$ submatrix  of $\gisig$:
\begin{equation}
\sqrt{\gisig_{2}} = \frac{1}{\sqrt{X}} \left[ \gisig_{2} + W   {\bf I}_2  \right],
\label{eqn:traceidentity}
\end{equation}
where
\begin{align}
\tX & \equiv\Bigl(\frac{\dot{f}}{b}+\mu\frac{g'}{\ta }\Bigr)^2-\Bigl(\frac{\dot{g}}{b}+\mu\frac{f'}{\ta }\Bigr)^2, \nonumber\\
W & \equiv \frac{\mu}{ab} \( \dot f g' - \dot g f' \),
\end{align}
and $\mu={\rm sgn}(\dot f g' - \dot g f')$.  

With Eq.~(\ref{eqn:traceidentity}),  traces of  $\gisig^n$ become
\begin{align}
\label{eqn:trace}
\tr{\sqrt{\gisig}} &= \sqrt{X} + \frac{2 g}{a r} , \\
\tr{\gisig} &= X  - 2 W + \frac{2 g^2}{a^2 r^2},\nonumber  \\
\tr{\gisig^{3/2}} &= X^{3/2} -3W\sqrt{X}  +  \frac{2 g^3}{a^3 r^3}, \nonumber\\
\tr{\gisig^2} &= X^2  - 2 W ( 2 X - W) + \frac{2 g^4}{a^4 r^4},  \nonumber
\end{align}
and the potential is given by
\begin{equation}
{\frac{\mathcal U}{4}} = P_0\left( \frac{g}{\ta r} \right) + \sqrt{\tX}P_1\left( \frac{g}{\ta r} \right)
+  W P_2\left( \frac{g}{\ta r} \right),
\label{eqn:genpot}
\end{equation}
where the $P_n$ polynomials are
\begin{align}
P_0(x) &= - 12 - 2 x(x-6) - 12(x-1)(x-2)\alpha_3 
\nonumber\\&\qquad -24(x-1)^2\alpha_4 ,\nonumber\\
P_1(x) &= 2 (3 -2 x)  +  6(x-1)(x-3)\alpha_3 +   24(x-1)^2 \alpha_4,\nonumber\\
P_2(x) &= -2 + 12 (x-1) \alpha_3 - 24(x-1)^2 \alpha_4.
\end{align}
Varying the action with respect to $f$ and $g$ yields the \stucky\ field equations
\begin{align}
\label{eqn:eomf}
&\partial_t\Biggl[\frac{\ta ^3r^2}{\sqrt{\tX}}\Bigl(\frac{\dot{f}}{b}+\mu\frac{g'}{\ta }\Bigr)P_1\parg + \mu\ta ^2 r^2 g' P_2\parg\Biggr] \\
&- \partial_r\Biggl[\frac{\ta ^2br^2}{ \sqrt{\tX}}
\Bigl(\mu\frac{\dot{g}}{b}+\frac{f'}{\ta }\Bigr)P_1\parg
+  \mu\ta ^2r^2 \dot{g}P_2\parg\Biggr]=0,
\nonumber
\end{align}
and
\begin{align}
-& \partial_t\Biggl[\frac{\ta ^3 r^2}{\sqrt{{\tX}}}\Bigl(\frac{\dot{g}}{b}+\mu\frac{f'}{\ta }\Bigr)
P_1\parg
+\mu \ta ^2r^2 f' P_2\parg\Biggr]\nonumber\\
& +\partial_r\Biggl[\frac{\ta ^2br^2}{\sqrt{\tX}}\Bigl(\mu\frac{\dot{f}}{b}+\frac{g'}{\ta }\Bigr)P_1\parg + \mu\ta ^2 r^2\dot{f}P_2\parg \Biggr]\notag\\
&=\ta^2 b r \left[ P_0' +  \sqrt{\tX} P_1'+ W P_2'\right] ,
\label{eqn:eomg}
\end{align}
where $P_n'(x)\equiv dP_n/dx = a r  \partial P/\partial g$.  
By inspection, we find that a solution to the $f$ equation of motion, Eq.~(\ref{eqn:eomf}),
is given by $P_1(x_0)=0$, or 
\begin{equation}
x_0 = \frac{ 1 + 6\alpha_3 + 12\alpha_4 \pm \sqrt{ 1+ 3\alpha_3 + 9\alpha_3^2 - 12 \alpha_4}}{3 (\alpha_3+4\alpha_4)}, \label{gsol}
\end{equation}
 and hence $g = x_0 \ta r$.
Note that if $\alpha_3=\alpha_4=0$, $P_1(x)$ becomes linear and $g = 3\ta r/2$ is the solution.

The equation of motion for $g$ evaluated on the solution provides a constraint on $f$ 
\begin{align}
\sqrt{\tX} P_1' &=\left( \frac{2P_2}{x_0}-P_2' \right) W  - P_0',
\label{eqn:Xfeom}
\end{align}
where the $P_n$ functions are evaluated at $x_0$ and we have 
used the fact that
\begin{align}
W &= \frac{\mu}{b}  \Bigl(\dot{f}+\frac{\ta '}{\ta }r\dot{f}-\frac{\dot{\ta }}{\ta }rf'\Bigr)x_0 .
\end{align}

An explicit solution for $f$ is not required for the computation of the stress energy tensor. That is, the physical background
solution does not depend on the choice of solution for $f$ and in particular is independent
of the spatial and temporal integration constants that are introduced in solving for $f$.
 After straightforward but tedious algebra, we find that its nonzero components are:
\begin{align}
T_{00}^{\mathcal{K}}
&=\frac{1}{2} P_0(x_0)b^2 ,
\nonumber\\
T_{rr}^{\mathcal{K}} 
&=-\frac{1}{2} P_0(x_0)\ta ^2 ,
\nonumber\\
T_{\theta\theta}^{\mathcal{K}}
= \frac{T_{\phi\phi}^\mathcal{K}}{\sin^2\theta}
&=-\frac{1}{2} P_0(x_0)\ta ^2r^2 .
\end{align}
The $T_{00}^{\mathcal{K}}$ and $T_{rr}^{\mathcal{K}}$ pieces can be easily checked from Eq.~(\ref{eqn:genpot}) by
direct variation with respect to $g^{tt}$ and $g^{rr}$, noting that the polynomial pieces come from the angular metric.
The angular pieces can be similarly analyzed by separately tracking the equal $\theta$ and $\phi$ contributions to $2 (g/ar)^n$ terms in the traces of Eq.~(\ref{eqn:trace}).  Their separate variations can then be reduced with Eq.~(\ref{eqn:Xfeom}). 

Hence, the effective energy density and pressure  are
\begin{equation}
({m^2}{M_{\rm pl}^2} )T^{\mu (\mathcal{K})}_{\;\,\nu} = \left(
      \begin{array}{cccc}
        -\rho_\mathcal{K} & 0 & 0 & 0 \\
        0 & p_\mathcal{K}  & 0 & 0 \\
        0 & 0 & p_\mathcal{K}  & 0 \\
        0 & 0 & 0 & p_\mathcal{K} 
      \end{array} \right), 
  \end{equation}
  where 
 \begin{equation}
 \rho_\mathcal{K} = -p_\mathcal{K} =\frac{1}{2} m^2 M_{\rm pl}^2 P_{0}(x_0).
 \end{equation}
  
This shows that a cosmological constant type solution exists for general isotropic metrics.   Conversely, the modified Einstein  equation for arbitrary spherically symmetric distributions of matter becomes
the ordinary Einstein equation plus a cosmological constant on this solution.

For example, the spatially flat FRW space-time is a subset where
$\ta(r,t)=a(t)$ is the scale factor,  $b(r,t)=1$ and the modified Einstein equation (\ref{eqn:modeinstein}) just becomes the usual Friedmann equation
\begin{equation}
\left( \frac{\dot a}{a} \right)^2 = \frac{1}{3 M_{\rm pl}^2}(\rho_\mathcal{K} + \rho_m).
\end{equation}
FRW space-times with spatial curvature $K\ne0$ are also included with
\begin{equation}
a(r,t) \rightarrow \frac{a(t)}{\sqrt{1+K r^2/4}}
\end{equation}
in isotropic coordinates.
  Note that for the FRW metric, this
 solution applies for radiation and matter domination as well as for a self-accelerated epoch where the massive graviton
 itself provides the cosmological constant-like dark energy.
It also allows for arbitrary isotropic perturbations around the FRW metric with 
\begin{align} 
\ta^2(r,t) = & a^2(t)[1+2 \Phi(r,t)] ,\nonumber\\
b^2(r,t) = & [1+2\Psi(r,t)].
\end{align}
Thus the solution remains of the cosmological constant type for arbitrary
spherically symmetric matter distributions.  Furthermore, the matter only sees the effects of the mass term as a cosmological constant with no direct coupling to the 
\stucky\  fields on the exact solution.

It is straightforward to verify that our class of solutions subsumes several particular solutions that have previously appeared in the literature: 
in vacuum ($T^{(m)}_{\mu\nu}$=0), it recovers exactly the static Schwarzschild-de Sitter solution
from \cite{Koyama:2011xz,Koyama:2011yg} and \cite{Nieuwenhuizen:2011sq}; it also reproduces the decoupling limit solution in \cite{deRham:2010tw,Koyama:2011wx} as well as
the open universe solution reported in \cite{Gumrukcuoglu:2011ew}.  Reduction to these particular solutions is effectively made through a choice of $f(t,0)$, thus demonstrating that those solutions are not unique.  Our solution is also  similar to another Schwarzschild-de Sitter
solution  \cite{Berezhiani:2011mt}, as well as the  vierbein formulation solution of \cite{Chamseddine:2011bu}.

\section{Discussion}

The solution we have found is a perfect analog for a cosmological constant. Because the solution exists
for any isotropic distribution of matter, it recovers static solutions like Schwarzschild-de Sitter in vacuum and generalizes them to dynamical cases such as the FRW cosmology. In each of these cases, the presence of other isotropic sources of stress-energy does not alter the cosmological constant-like behavior of massive gravity.
Hence, we can have a truly self-accelerating gravitational background that coexists peacefully with a standard
cosmological history; the self acceleration begins in precisely the same manner as cosmological-constant-driven acceleration would begin,
only here the size of the apparent cosmological constant is set by the graviton mass and the other free parameters of the theory ($\alpha_3$
and $\alpha_4$). Moreover, the extra gravitational degrees of freedom present in this theory do not appear to couple to radial matter perturbations. 
In practice, this means that radial matter perturbations will feel only the ordinary gravitational attraction as in general relativity. This is
 in contrast with the enhanced gravitational force felt by matter perturbations around other solutions of this theory. Likewise the \stucky-driven self-accelerating background physics will not respond
 to spherically symmetric matter perturbations either.

In this paper, we have restricted ourselves to isotropic situations. Note that although we have assumed our \stucky\ fields to be in a radially symmetric configuration, their effective center in space disappears in their effective stress energy,
which is homogeneous and isotropic. Indeed,
we can recover fully homogeneous background solutions supported by the \stucky\ fields. This suggests
that perhaps even more general inhomogeneity in the matter  fields may not drive inhomogeneity in the observable effective stress energy of the \stucky\ fields.

The obvious next step in assessing the solution we have found is to attempt to study perturbations around it.  Naively speaking, 
the solution we have found gives us reason for concern, because the part of the action that we might expect to generate kinetic terms for these fluctuations
appears to vanish on our solution, since $P_1(x_0)=0$.  At first, this appears to confirm the findings of  \cite{Gumrukcuoglu:2011zh}, who find vanishing kinetic terms for perturbations of the \stucky\ fields
around their open Universe solution \cite{Gumrukcuoglu:2011ew}, which is a member of our class of solutions.
 However, our class also includes the solutions found in \cite{Koyama:2011wx}, who studied vector and scalar perturbations in the decoupling limit of their solution and found non-vanishing kinetic terms for the scalar perturbations as well as either ghost-like or vanishing kinetic terms for the vector perturbations. 
 Similarly, the closely related decoupling limit solution of  \cite{deRham:2010tw} also has non-vanishing kinetic terms for scalar perturbations. 
  In light of these considerations, a careful study of general perturbations to our class of solutions will be an important area for future work.

\smallskip{\em Acknowledgments.--}  We thank P. Adshead, C. de Rham, G. Dvali, K. Koyama, R. Rosen, and A. Tolley for helpful discussions. 
PG was supported by the National Research Fund Luxembourg through grant BFR08-024. WH was supported by Kavli Institute for Cosmological Physics at the University of Chicago through grants NSF PHY-0114422 and NSF PHY-0551142  and an endowment from the Kavli Foundation and its founder Fred Kavli and by the David and Lucile Packard Foundation. MW was supported, and WH was additionally supported, by U.S.~Dept.\ of Energy contract DE-FG02-90ER-40560. 
 
\bibliography{paperbib}

\end{document}